# A topological Dirac-vortex parametric phonon laser

Xiang Xi[†], Jingwen Ma[†] and Xiankai Sun[*]

*Department of Electronic Engineering, The Chinese University of Hong Kong, Shatin, New Territories, Hong Kong*

[†]*These authors contributed equally to this work*

[*]*Corresponding author: xksun@cuhk.edu.hk*

**Nonlinear topological photonic and phononic systems have recently aroused intense interests in exploring new phenomena that have no counterparts in electronic systems. The squeezed bosonic interaction in these systems is particularly interesting, because it can modify the vacuum fluctuations of topological states, drive them into instabilities, and lead to topological parametric lasers. However, these phenomena remain experimentally elusive because of limited nonlinearities in most existing topological bosonic systems. Here, we experimentally realized topological parametric lasers based on nonlinear nanoelectromechanical Dirac-vortex cavities with strong squeezed interaction. Specifically, we parametrically drove the Dirac-vortex cavities to provide phase-sensitive amplification for topological phonons, and observed phonon lasing above the threshold. Additionally, we confirmed that the lasing frequency is robust against fabrication disorders and that the free spectral range defies the universal inverse scaling law with increased cavity size, which benefit the realization of large-area single-mode lasers. Our results represent an important advance in experimental investigations of topological physics with large bosonic nonlinearities and parametric gain.**

The discovery of topologically nontrivial electronic systems is a significant milestone in condensed-matter physics[1]. While many topological phases (e.g., quantum Hall, quantum spin Hall, and quantum valley Hall) are based on noninteracting electronic systems, there are intense research activities in exploring topological phases with strong interactions[2]. One important example is the topological superconductors, where the nontrivial topology and the fermionic squeezed interaction can jointly lead to the emergence of Majorana fermions[3]. Recently, topological physics has been extended to the realms of optics[4, 5] and acoustics[6, 7]. The bosonic nature of photons and phonons has aroused intense research interests in investigating topological physics in nonlinear optical or acoustic systems[8]. The third-order nonlinearities in these systems have enabled various important functions or phenomena, including topological third-harmonic generation[9, 10], topological entangled photon-pair generation[11, 12], topological solitons[13, 14], and tunable topological states[15, 16]. On the other hand, the second-order nonlinearities are also important, because they can provide strong bosonic squeezed interactions between photons or phonons that will unlock new topological phases and functionalities fundamentally different from their counterparts in the electronic domain[17]. For instance, parametric lasers (or parametric oscillators) are one of the most important results of bosonic squeezed interactions. They exhibit much lower noises than conventional semiconductor lasers and are widely used as coherent optical/microwave sources for fundamental research or in practical applications. While topological theories have recently been applied to conventional semiconductor lasers[18, 19], topological parametric lasers[20] remain experimentally elusive. This is because of negligible second-order nonlinearities in most existing topological photonic and phononic systems[4-7], which cannot provide sufficient squeezed interactions.

Dirac-vortex states are a new type of topological states recently discovered in two-dimensional photonic[21, 22] and phononic[23-25] systems. They are described by the Jackiw–Rossi model[26], and are mathematically identical to the Majorana fermions in topological superconductors. Unlike conventional resonant states in topologically trivial systems, Dirac-vortex states with a scalable modal area $S$ can have an unprecedentedly large free spectral range (FSR) that defies the universal inverse relationship FSR ~ $1/S$. Besides, their resonant frequencies are topologically pinned to the bulk Dirac frequencies and remain nearly constant with respect to $S$. These properties make Dirac-vortex states an ideal candidate for single-mode large-area topological lasers. However, previous experimental demonstrations of Dirac-vortex states are limited to passive photonic and phononic systems. Despite intense research activities in topological lasers[18, 19], lasing from the topological Dirac-vortex states remains to be obtained.

Here, we experimentally demonstrated a topological parametric laser from the Dirac-vortex states on a nonlinear nanoelectromechanical platform. Such nanoelectromechanical systems exhibit strong second- and third-order nonlinearities leading to various phenomena, such as mode cooling[27], vacuum-state squeezing[28], synchronized oscillating[29], and phonon lasing[30]. We harnessed the strong squeezed interaction on this platform for parametric amplification of topological phonons. We observed phonon lasing above the threshold associated with linewidth-narrowing effect. Additionally, we confirmed that the lasing frequency is robust against fabrication disorders and that the free spectral range defies the universal inverse scaling law with increased cavity size. Our results not only provide a robust coherent phonon source for topological integrated phononic circuits, but also represent an important leap forward in experimental investigation of the interplay between topology and nonlinearities.

Figure 1a is a conceptual illustration of a topological Dirac-vortex parametric phonon laser. A two-dimensional array of suspended silicon nitride membranes with thickness of 150 nm was formed by removing the underlying sacrificial oxide in parts. A 40-nm-thick aluminum layer was deposited on top of silicon nitride to serve as the signal electrode. The heavily doped silicon substrate served as the electrical ground. Figure 1b is a scanning electron microscope image of the fabricated nanoelectromechanical crystal with lattice constant $l_0 = 20.8$ μm. Each unit cell contains six nanoelectromechanical membranes, which can be classified into two groups as colored in red and blue. Their geometries are controlled by the relative positions of the small holes (600 nm in diameter) etched in the silicon nitride layer $(r_1, r_2) = (r_0 - \delta_0\cos\theta + \delta_0\sin\theta, r_0 - \delta_0\cos\theta - (\delta_0\sin\theta)/2)$ and $(r_3, r_4) = (r_0 + \delta_0\cos\theta + \delta_0\sin\theta, r_0 + \delta_0\cos\theta - (\delta_0\sin\theta)/2)$ with $r_0 = 5.4$ μm. The effective bulk Hamiltonian of the crystal is $H(\mathbf{k}) = v_\mathrm{D} \cdot (\sigma_x k_x + \sigma_y k_y) - \Delta_0/2 \cdot \sigma_z(\tau_x \cos\theta + \tau_y \sin\theta)$, where $\sigma_x$, $\sigma_y$, $\sigma_z$, $\tau_x$, and $\tau_y$ are the Pauli matrices, $\mathbf{k} = (k_x, k_y)$ is the wave vector, $\Delta_0$ is the bulk bandgap proportional to $\delta_0$, and $v_\mathrm{D}$ is the effective Fermi velocity (see Supplementary Sec. 1). As shown in Fig. 1c, the device has spatially varying parameters $\delta_0(\mathbf{r}) = \delta_{\max}\tanh(|\mathbf{r}|/R_0)$ and $\theta(\mathbf{r}) = \arg(\mathbf{r})$, with $\delta_{\max} = 300$ nm and $R_0$ controlling the spatially varying bulk bandgap along the radial direction. Such a configuration is mathematically identical to the Jackiw–Rossi model[26] and can support the Dirac-vortex states[25]. Theoretically, the modal profile of the Dirac-vortex state is $f_0(\mathbf{r}) = g_0(|\mathbf{r}|) \cdot |\psi_0\rangle$, which is composed of an envelope function $g_0(|\mathbf{r}|)$ controlling the modal area $S$ and a

periodic Bloch mode $|\psi_0\rangle$ controlling the modal profile within each unit cell (see Supplementary Sec. 1). Figure 1d shows the simulated intensity modal profile $|f_0(\mathbf{r})|^2$ of the Dirac-vortex state with $R_0/l_0 = 0.5$. Figure 1e shows the experimental mechanical intensity spectra for the device directly actuated in the frequency range of 41–49 MHz. In the bulk bandgap region, we can find the topological Dirac-vortex state with a resonant frequency $\omega_0/2\pi = 45.238$ MHz and a quality factor ~2,874 measured at room temperature. The damping rate of the Dirac-vortex state is $\gamma/2\pi = 7.9$ kHz.

To provide strong squeezed interaction for the Dirac-vortex state, a combination of a d.c. bias voltage $V_0$ and an a.c. parametric pump voltage $V_{\text{pump}}\cos(2\Omega t)$ was applied across the signal and ground electrodes to exert an electrocapacitive force

$$F = \frac{\varepsilon\left[V_0 + V_{\text{pump}}\cos(2\Omega t)\right]^2}{2\left[d - W(\mathbf{r},t)\right]^2} \tag{1}$$

on the nanoelectromechanical membranes (Fig. 1a), where $\Omega$ is near the resonant frequency $\omega_0$ of the Dirac-vortex state with a frequency detuning $\Delta = \Omega - \omega_0$, $\varepsilon$ is the effective permittivity, $d$ is the distance between the silicon substrate and the aluminum layer deposited on silicon nitride, and $W(\mathbf{r}, t) = a(t)f_0(\mathbf{r})e^{j\Omega t} + $ h.c. is the vibrational displacement with $a(t)$ and $f_0(\mathbf{r})$ describing the temporal dynamics and spatial modal profile of the Dirac-vortex state, respectively. Being dependent on $W(\mathbf{r}, t)$, the electrocapacitive force $F$ can modify the intrinsic vibrational dynamics of the Dirac-vortex state, leading to squeezed interaction. The temporal evolution of $a(t)$ is governed by a second-quantized Hamiltonian

$$H = \left(\Delta + \alpha V_{\text{pump}}^2/2 - \beta\hat{a}^\dagger\hat{a}\right)\hat{a}^\dagger\hat{a} + \frac{\alpha V_0 V_{\text{pump}}}{2}\left(\hat{a}^\dagger\hat{a}^\dagger + \hat{a}\hat{a}\right), \tag{2}$$

where $\hat{a}$ and $\hat{a}^\dagger$ are respectively the annihilation and creation operators for the topological Dirac-vortex state, $\beta$ is the Kerr nonlinear coefficient, and $\alpha$ is the electromechanical tuning coefficient (see Supplementary Sec. 2). The first term in Eq. (2) represents the Dirac-vortex mode whose frequency is shifted by the external electric field and the third-order nonlinearity, while the second term represents the single-mode squeezed interaction. The measured vibration quadrature in Fig. 1f shows that the parametric pump $V_{\text{pump}}\cos(2\Omega t)$ can squeeze the Dirac-vortex state (Extended Data Fig. 1) and provide phase-sensitive amplification (Extended Data Fig. 2).

When the parametric amplification is sufficiently large to overcome the dissipation, the squeezed interaction can drive the zero-solution $a(t) = 0$ of the system into instability. Consequently, any fluctuations can cause exponential growth of $a(t)$ and finally the parametric phonon lasing. The steady-state power of the parametric laser is determined by the frequency detuning $\Delta$ and pump voltage $V_{pump}$:

$$|a(t)|^2 = \frac{\Delta + \alpha V_{pump}^2/2 + \alpha V_0 \sqrt{V_{pump}^2 - V_{th}^2}}{\beta}, \tag{3}$$

where $V_{th} = \gamma/\alpha V_0$ is the minimal threshold pump voltage when the frequency detuning satisfies $\Delta = -\alpha V_{th}^2/2$ (see Supplementary Sec. 3). We characterized the lasing behavior of the device with $R_0/l_0 = 0.5$ by applying a coherent parametric pump. Figure 2a plots the measured power spectral density (PSD) of phonon laser emission with $V_{pump} = 11$ V, which shows that the peak frequencies of the lasing spectra are always at $\Omega$, i.e., half of the pump frequency $2\Omega$. Figure 2b plots the measured peak PSD as a function of the pump frequency $2\Omega$ and pump voltage $V_{pump}$, which agrees well with the theoretical prediction of Eq. (3) (see Extended Data Fig. 3 and Supplementary Sec. 3). Figure 2c plots the peak PSD as a function of the pump voltage $V_{pump}$ at a fixed pump frequency $2\Omega/2\pi = 90.44$ MHz. Figure 2d plots the peak PSD as a function of the pump frequency at a fixed pump voltage $V_{pump} = 8.9$ V, which shows a linear relationship between $|a(t)|^2$ and the pump frequency $2\Omega$. By fitting the experimental results in Fig. 2c and 2d with the analytical solution in Eq. (3), we obtained the minimal threshold pump voltage $V_{th} = 2.09$ V, the Kerr nonlinear coefficient $\beta = 0.392$ kHz nm$^{-2}$, and the electromechanical tuning coefficient $\alpha = 0.40$ kHz V$^{-2}$. Note that for $2\Omega/2\pi > 90.477$ MHz the system enters a bistable region (gray shaded region in Fig. 2d), where the parametric laser emission cannot be obtained because the zero-solution $a(t) = 0$ is stable (see Supplementary Sec. 3).

While the existence of a threshold in the stimulated phonon emission is a typical characteristic of lasing, another fundamental characteristic of laser is spectral linewidth narrowing above the threshold. To demonstrate that, we applied an incoherent white-noise voltage centered at frequency $2\Omega/2\pi = 90.43$ MHz with a bandwidth of 20 kHz to parametrically pump the device. Figure 3a and 3b show the measured PSD under various pump voltage $V_{pump}$. Figure 3c shows the measured modal profile above the lasing threshold, which agrees well with the simulated result in Fig. 1d. Figure 3d shows the measured peak PSD of the device as a function of $V_{pump}$ under the incoherent pump, indicating a threshold pump voltage of 2.75 V. This threshold voltage is higher than that under a coherent pump (Fig. 2c) because of a larger pump bandwidth. Figure 3e shows that the linewidth fitted from the measured spectra in Fig. 3a reduces from 11.0 to 3.8 kHz when $V_{pump}$ increases to 4.45 V, which is a distinctive evidence of lasing. Figure 3f plots the fitted peak frequency of the lasing spectra as a function of the pump voltage. Below the threshold, the peak frequency increases rapidly, because the squeezed interaction changes the real part of the eigenfrequency of the Hamiltonian in Eq. (2) before it is sufficiently strong to drive the system into instability (see Supplementary Sec. 2). On the other hand, when the pump voltage is sufficiently large, it induces a redshift to the lasing frequency which is accompanied with slight broadening of the laser linewidth as shown in the light blue regions in Fig. 3e and 3f.

We further investigated the Dirac-vortex lasers with different $R_0$ (Fig. 4a). Figure 4b shows the experimental mechanical intensity spectra of the devices with different $R_0$ measured under direct actuation at frequencies from 42 to 48 MHz, confirming the existence of the Dirac-vortex states in the bulk bandgap region. Additionally, Fig. 4c shows that the resonant frequencies of devices with an identical design can vary due to fabrication-induced disorders. The relative frequency fluctuation $\delta\omega/\omega_0$ ($\delta\omega$: standard deviation of the resonant frequency) reduces from 0.16% to 0.045% as $R_0/l_0$ is increased from 0.5 to 4, which indicates that the Dirac-vortex states with a larger $R_0$ are more robust against the fabrication-induced disorders. Note that the frequency fluctuation due to fabrication-induced disorders is considerably smaller than typical nontopological nanomechanical devices[31]. Furthermore, we experimentally confirmed that all the devices with different $R_0$ values can support phonon lasing under an incoherent parametric pump, as shown by the measured PSD in Fig. 4d. Figure 4e shows the intensity modal profiles $|f_0(\mathbf{r})|^2$ measured above the lasing threshold along the $x$ direction as indicated by the white dashed arrow in Fig. 4a. Their envelopes (orange dashed lines in Fig. 4e) agree well with the analytical function $|g_0(|\mathbf{r}|)|^2 = [\cosh(|\mathbf{r}|/R_0)]^{-R_0/\varsigma}$, where the experimentally fitted $\varsigma$ was found to be $0.52l_0$. Figure 4f plots the devices' modal area $S = \pi D_{\text{eff}}^2/4$ as a function of $R_0$ (orange stars), where $D_{\text{eff}}$ is the effective mode diameter defined by the full width at half maximum of the fitted function $|g_0(|\mathbf{r}|)|^2$ in Fig. 4e. Meanwhile, Fig. 4f also shows the measured resonant frequencies of the desired Dirac-vortex state and its sideband states (purple dots). These results confirm that the FSR of the Dirac-vortex states defies the universal inverse relationship FSR ~ $1/S$, which indicates that the Dirac-vortex cavities are suitable for realizing large-area single-mode lasers.

In conclusion, we experimentally demonstrated a topological Dirac-vortex parametric phonon laser by harnessing the strong squeezed interactions in a nanoelectromechanical system. Our results will excite broad interests in the following aspects. First, our topological phononic system with strong squeezed interaction offers a practical experimental platform for investigating the interplay between topology and nonlinearity. It can be used for studying many important phenomena such as topological solitons, topological entangled phonon-pair generation, topological nonclassical states preparation, and topological synchronization and chaos. Second, the experimentally realized parametric gain offers a promising strategy to investigate the non-Hermitian topological physics, which is still experimentally elusive in the phononic domain. Third, considering that a similar nonlinear Hamiltonian has already been adopted for bit storage and flipping operations in nanoelectromechanical systems[32], our results can readily lead to topologically protected phononic computing technologies. Fourth, the fabrication robustness and electrical tunability of the Dirac-vortex cavities can be harnessed for realizing indistinguishable mechanical states that are desired in quantum entanglement and teleportation experiments[31]. Finally, our results demonstrating the first Dirac-vortex phonon lasers can readily be extended to realize single-mode large-area topological lasers in the optical domain.

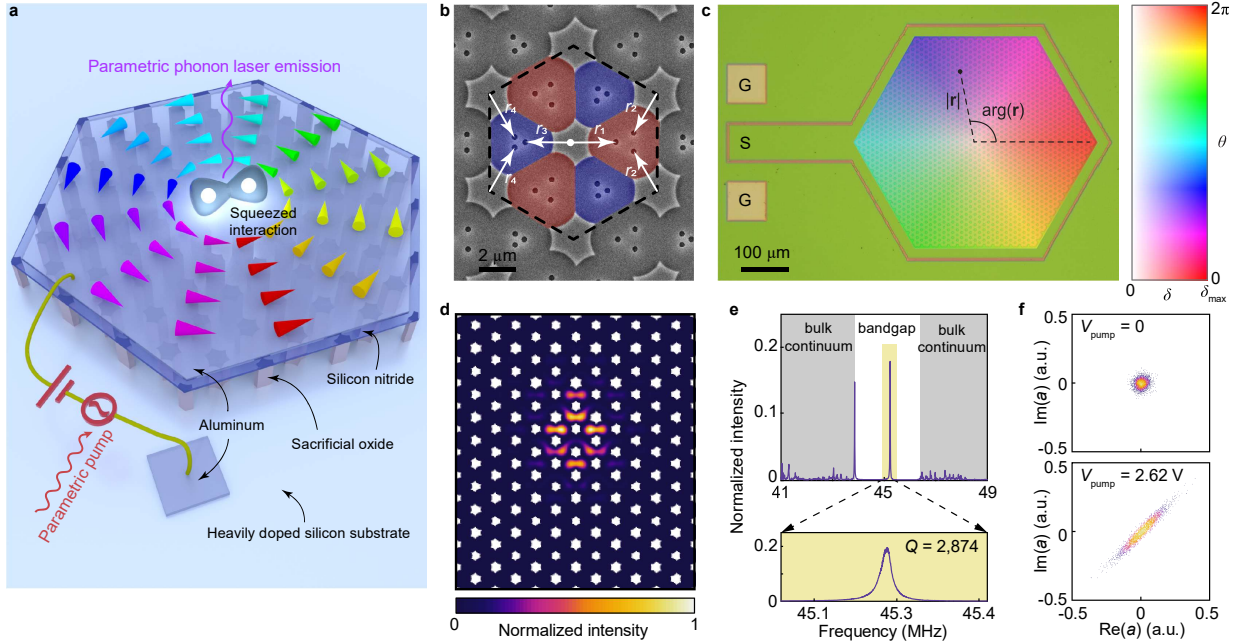

**Fig. 1 | Nanoelectromechanical Dirac-vortex states with squeezed interaction. a**, Conceptual illustration of a nanoelectromechanical Dirac-vortex parametric phonon laser. The device is based on a 2D array of suspended silicon nitride membranes, which encompasses a topological phase winding process rendered by the swirling cones. The heavily doped silicon substrate serves as the electrical ground, and the aluminum layer deposited on silicon nitride is connected to the signal electrode. Applying a combination of a d.c. bias voltage and an a.c. parametric pump voltage across the signal and ground electrodes leads to a strong squeezed interaction for the topological phonons near the center of the device, which enables a Dirac-vortex parametric phonon laser. **b**, Scanning electron microscope image of the nanoelectromechanical crystal (before aluminum deposition) showing a unit cell of the hexagonal lattice (black dashed hexagon). Each unit cell contains two groups of suspended membranes (colored in blue and red), whose geometries are determined by the positions of the etched holes ($r_1$, $r_2$) and ($r_3$, $r_4$). **c**, Optical microscope image of the device (before aluminum deposition). It is color-coded by the spatially varying parameters $\delta_0(\mathbf{r}) = \delta_{max}\tanh(|\mathbf{r}|/R_0)$ and $\theta(\mathbf{r}) = \arg(\mathbf{r})$ with $R_0/l_0 = 0.5$. S, signal electrode; G, ground electrode. **d**, Simulated intensity modal profile of the Dirac-vortex state. **e**, Measured mechanical intensity spectra of the device in **c**. The gray shaded regions correspond to the bulk continuum. The Dirac-vortex state lies in the bulk bandgap region with a resonant frequency $\omega_0/2\pi = 45.238$ MHz and a quality factor ~2,874. **f**, Measured vibration quadrature of the Dirac-vortex state without ($V_{pump} = 0$) and with ($V_{pump} = 2.62$ V, pump frequency $2\omega_0$) squeezed interaction.

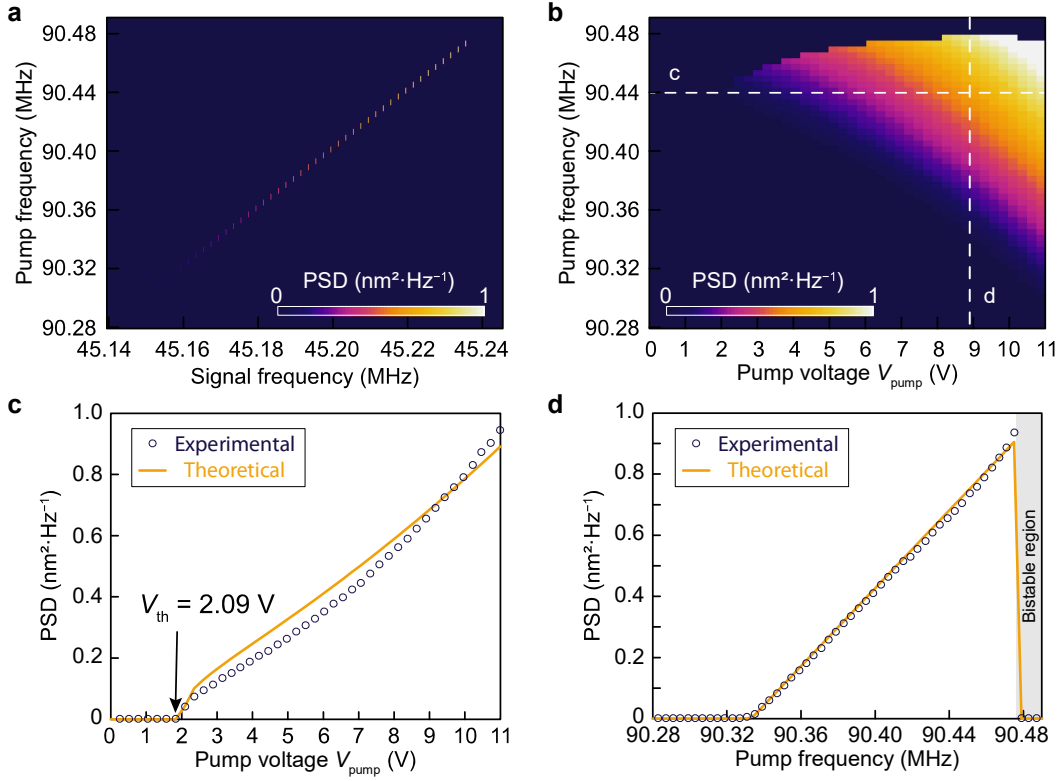

**Fig. 2 | Experimental demonstration of parametric phonon lasing from a Dirac-vortex state under a coherent pump. a**, Measured mechanical power spectral density (PSD) of the device with $R_0/l_0 = 0.5$ under the pump voltage $V_{pump} = 11$ V, showing that the lasing frequency is always half of the pump frequency. **b**, Measured peak PSD of the device with $R_0/l_0 = 0.5$ as a function of the pump frequency and pump voltage $V_{pump}$. **c**, Peak PSD as a function of $V_{pump}$ with the pump frequency fixed at 90.44 MHz (along the white dashed line c in **b**). **d**, Peak PSD as a function of the pump frequency with the pump voltage $V_{pump} = 8.9$ V (along the white dashed line d in **b**). The bistable region is marked in gray. In **c** and **d**, the dark blue circles and orange solid lines represent the measured data and theoretical fits, respectively.

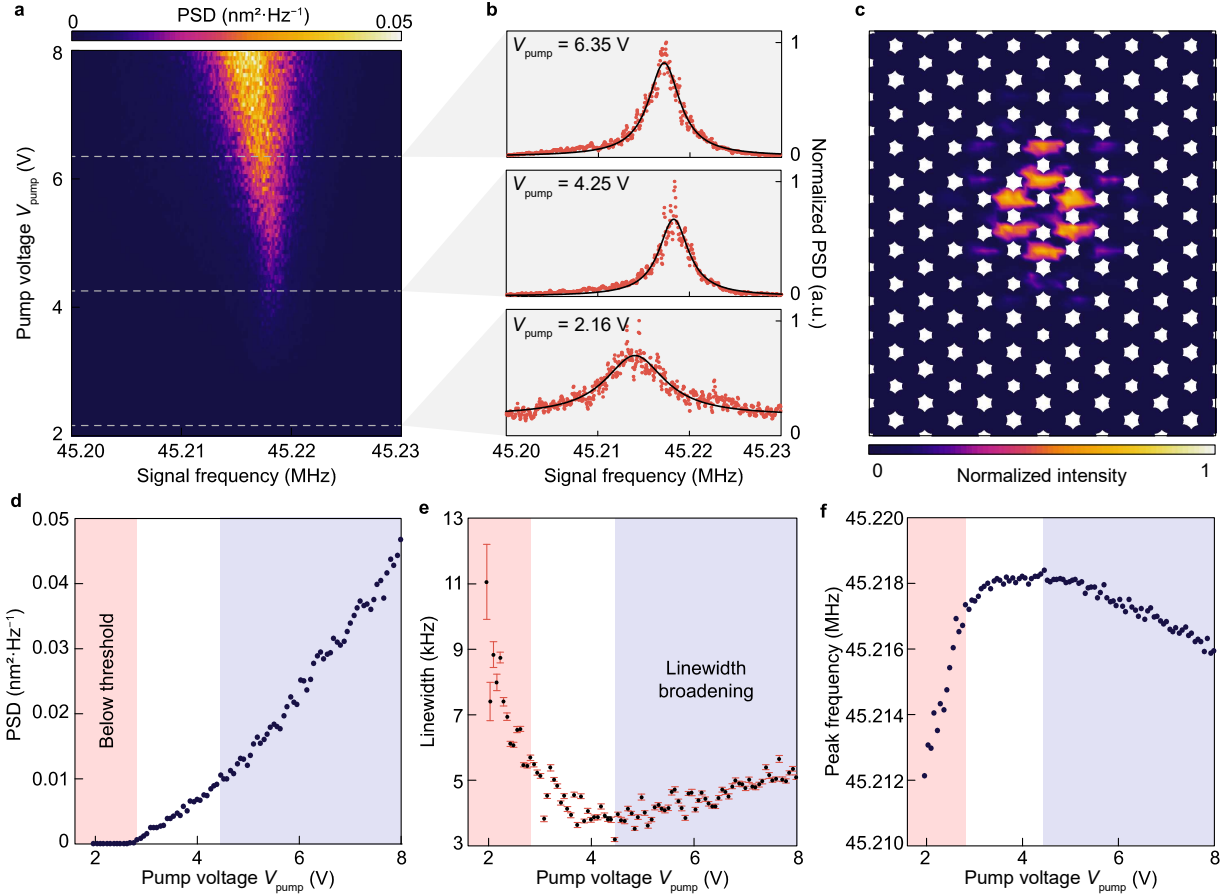

**Fig. 3 | Experimental demonstration of parametric phonon lasing from a Dirac-vortex state under an incoherent white-noise pump. a**, Measured mechanical PSD of the device with $R_0/l_0 = 0.5$ under different pump voltages. **b**, Measured normalized PSD under pump voltage $V_{pump} = 2.16$, 4.25, and 6.35 V. The solid black lines are Lorentzian fits of the measured data. **c**, Measured intensity modal profile of the Dirac-vortex state above the lasing threshold. **d**, Measured peak PSD of the device as a function of $V_{pump}$, showing a threshold pump voltage of 2.75 V. **e**, Fitted linewidth of the measured PSD as a function of $V_{pump}$. The error bars represent standard deviation during the linewidth fitting. **f**, Measured peak frequency as a function of $V_{pump}$. In **d**–**f**, the region below the lasing threshold is marked in pink. In **e** and **f**, the region of linewidth broadening is marked in light blue.

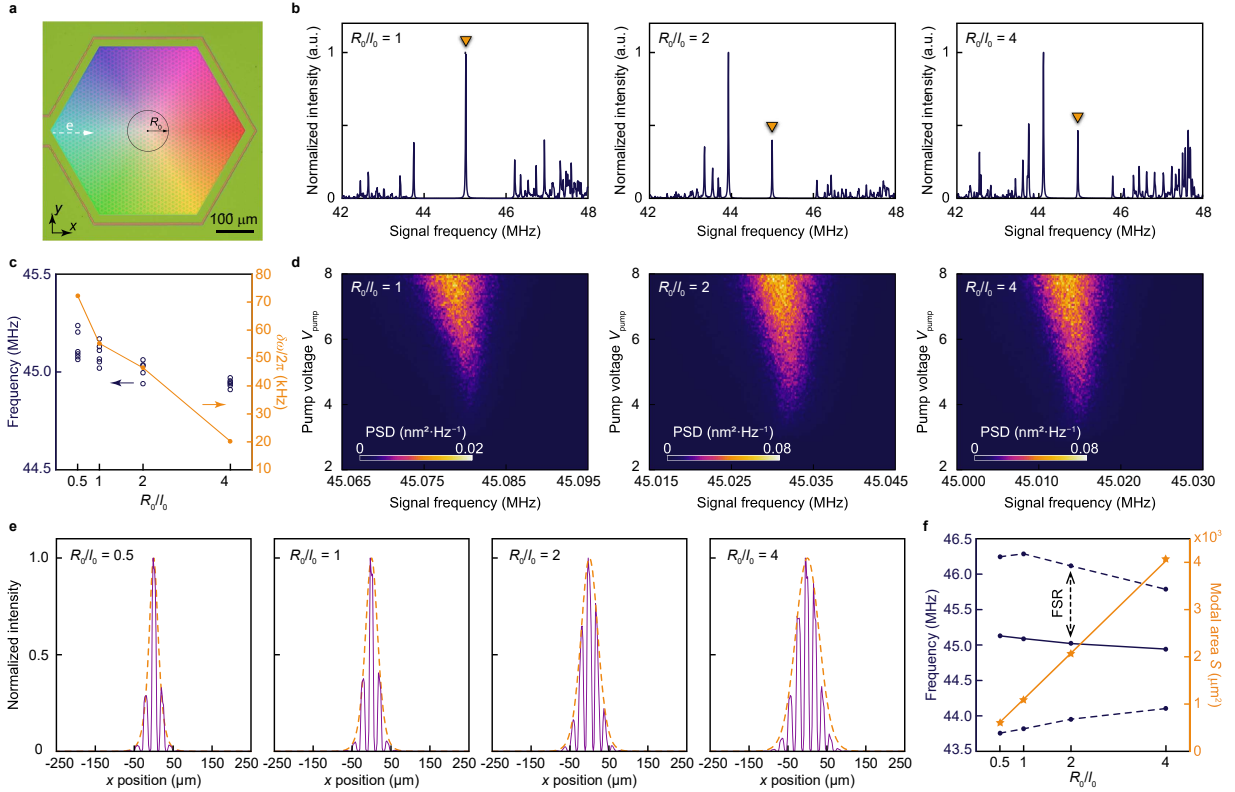

**Fig. 4 | Experimental results of Dirac-vortex parametric phonon lasers with different modal areas. a**, Optical microscope image of a fabricated device. The modal area $S$ of the Dirac-vortex state is controlled by the parameter $R_0$. **b**, Measured mechanical intensity spectra of the devices with different $R_0$ values, showing the existence of the Dirac-vortex state in the bulk bandgap region of all these devices. **c**, Measured resonant frequencies of the Dirac-vortex states with different $R_0/l_0$. The orange line plots the frequency fluctuation $\delta\omega$ measured from multiple devices with an identical design. We obtained $\delta\omega/\omega_0$ = 0.16%, 0.12%, 0.1%, and 0.045% for $R_0/l_0$ = 0.5, 1, 2, and 4, respectively. **d**, Measured mechanical PSD of the devices with $R_0/l_0$ = 1, 2, and 4 under an incoherent white-noise parametric pump with a frequency bandwidth of 20 kHz. **e**, Measured lasing intensity modal profiles $|f_0(\mathbf{r})|^2$ (purple solid lines) along the $x$ direction as indicated by the white dashed arrow in **a** from the devices with different $R_0$ values. The orange dashed lines plot the fitted envelope function $|g_0(|\mathbf{r}|)|^2$. **f**, Measured resonant frequencies (purple dots) of the Dirac-vortex state and its sidebands and the fitted modal area $S$ (orange stars) of the Dirac-vortex state from the devices with $R_0/l_0$ = 0.5, 1, 2, and 4.

13